\documentclass[aps,jmp,amsmath,amssymb,reprint]{revtex4-1}
\usepackage{graphicx}% Include figure files
\usepackage{dcolumn}% Align table columns on decimal point
\usepackage{bm}% bold math
\usepackage{graphicx}
\usepackage{subfigure}
\usepackage{physics}
\usepackage[english]{babel}
\usepackage[usenames, dvipsnames]{color}
\usepackage{hyperref}
\usepackage[normalem]{ulem}
\usepackage[utf8]{inputenc}
\usepackage{romannum}
\usepackage{float}
\begin{document}
\bibliographystyle{apsrev.bst}

\pagenumbering{arabic}
\renewcommand{\figurename}{FIG.}
\def\tablename{TABLE}

\title{Ultrahigh-performance superlattice mid-infrared nBn photodetectors at high operating temperatures}

%\title{Proposal for high performance mid-wavelength infrared ternary superlattice nBn photodetectors}

%\title{Efficient ultrahigh-performance nBn mid-infrared photodetectors}

%\title{Theoretical prediction of ultrahigh-performance superlattice nBn mid-infrared photodetectors}

%\title{Highly efficient mid-infrared superlattice nBn photodetectors for high operating temperatures }

%\title{High-temperature mid-infrared superlattice nBn photodetector based on InAsSb ternary alloy material system }

%\title{Mid-wavelength infrared high operating temperature nBn photodetectors based on InAsSb alloy material system}

%\title{Ultrahigh-responsivity superlattice mid-infrared nBn photodetectors based on InAsSb ternary absorber for high operating temperatures}

\author{Rohit Kumar$^1$}
\author{Bhaskaran Muralidharan$^1$}%
\thanks{corresponding author: bm@ee.iitb.ac.in}
%\homepage{\:bm@ee.iitb.ac.in}
\affiliation{$^1$Department of Electrical Engineering, Indian Institute of Technology Bombay, Powai, Mumbai-400076, India}
\date{\today}
\begin{abstract}
While advancing a physics-based comprehensive photodetector-simulation model, we propose a novel device design of the mid-wavelength infrared nBn photodetectors by exploiting the inherit flexibility of the InAs\textsubscript{1-x}Sb\textsubscript{x} ternary alloy material system. %Using the finite difference method in conjunction with the linear interpolation technique, we solve the Poisson and the continuity equations for the carriers while taking into account the temperature, doping, and structural dependence of the ternary alloy material system. 
To further explicate the physics of such photodetectors, we 
calculate several crucial transport and optoelectronic parameters, including the dark current density, absorption coefficient, responsivity, and the quantum efficiency of nBn photodetectors. A remarkable maximum efficiency of 57.39\% is achieved at room temperature at a bias of -0.25 V, coupled with a radiation power density of 50 mW/$cm^{2}$. The proposed structure features a maximum quantum efficiency of 44.18\% and 37.87\% at 60\% and 70\% of the $\lambda_c$, respectively. Furthermore, a maximum responsivity of 0.9257 A/W is shown within the mid-wavelength infrared spectrum. 
Through our comprehensive analysis, we also demonstrate that our proposed device design effectively reduces the dark current density by confining the electric field inside the barrier while preserving a superior level of quantum efficiency, and the current in such detectors is diffusion-limited. %Hence, the generation-recombination and tunneling currents do not limit the high performance of the nBn architecture. 
Insights uncovered here could be of broad interest to critically evaluate the potential of the nBn structures for mid-wavelength infrared photodetectors.
\end{abstract}
\maketitle

\begin{figure*}[!t]
	\centering
	\subfigure[]{\includegraphics[height=0.35\textwidth,width=0.55\textwidth]{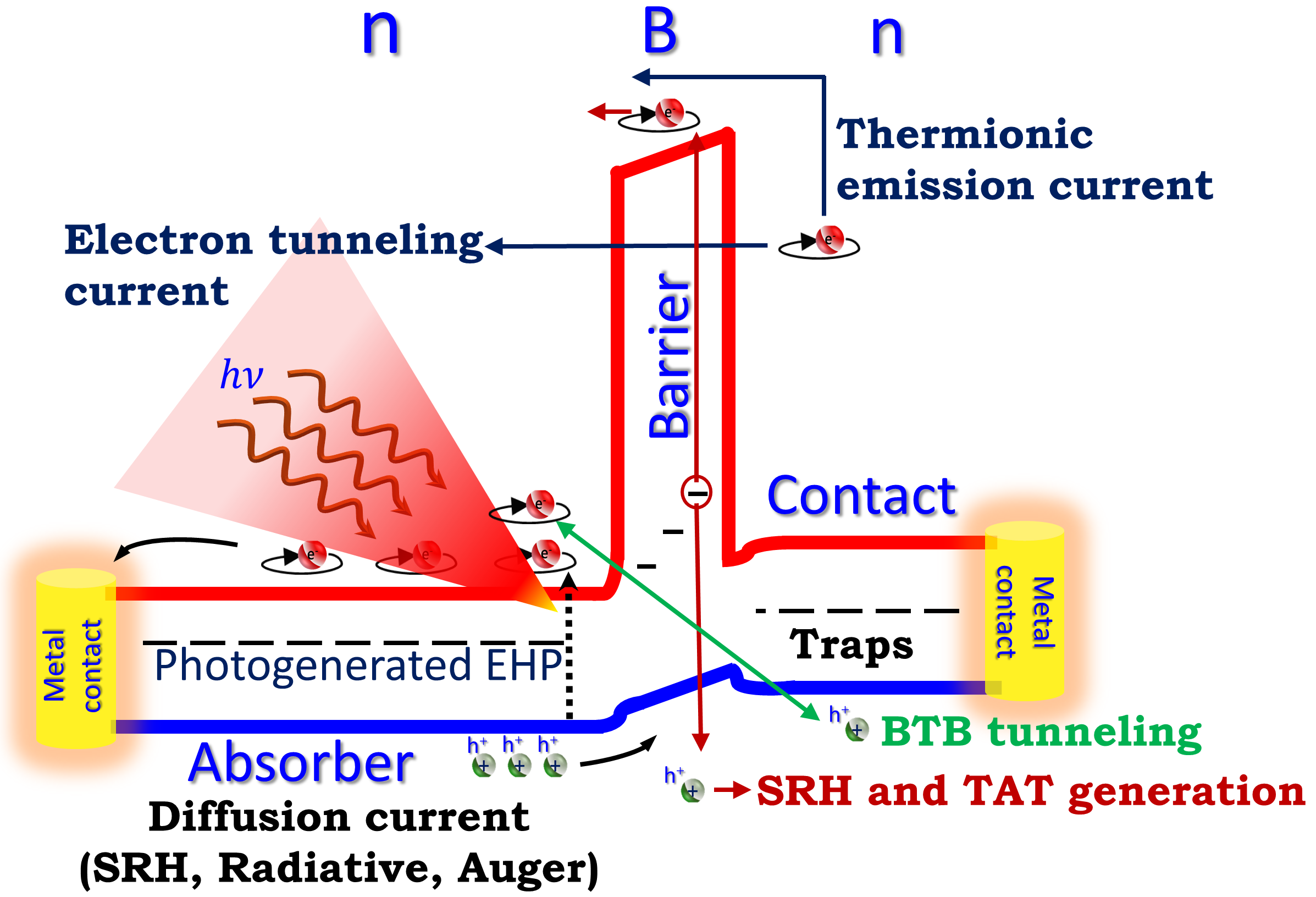}}\label{schematic}
   	\quad
	\subfigure[]{\includegraphics[height=0.34\textwidth,width=0.37\textwidth]{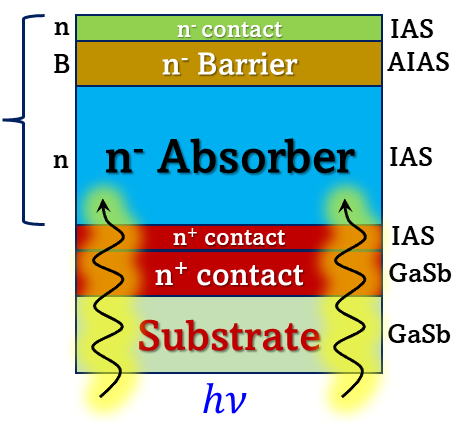}}\label{layout}
	\quad
	\caption{Preliminaries for the proposed MWIR nBn photodetector (a) schematic illustration of the carrier transport in the nBn photodetector (b) layout of the considered nBn photodetector with IAS as the AL and the CL, and AIAS as the BL that hinders the movement of electrons.	}
	\label{preliminaries}
\end{figure*}

\indent Mid-wavelength infrared (MWIR) photodetectors \cite{soibel2014room,kumar2023advancing,soibel2019mid,gautam2012high,wu2020high} are in high demand for a wide range of civilian, military, and space applications, including environmental monitoring, chemical sensing, medical diagnostics, and infrared (IR) imaging. This is because they meet high operating temperature requirements, exhibit better performance, and have numerous advantages over other photodetector candidates \cite{martyniuk2013SPIE,martyniuk2014new,klipstein2008xbn,rogalski2020inassb}. The design of such a high performance MWIR photodetector relies on achieving a balance between low dark current and high quantum efficiency. A majority of photodiodes in the market today are p-n junction photodiodes made from conventional materials like HgCdTe (MCT) \cite{itsuno2012mid}, InGaAs \cite{martinelli19882} and InSb \cite{shimatani2020high}, etc., which are plagued by space-charge generation-recombination (G-R) dark currents that significantly restrict their efficacy for applications demanding high sensitivity at lower temperatures \cite{reine2014numerical}. In order to prevent excessive dark currents, these devices typically need to be cooled down to cryogenic temperatures. \\
\indent The nBn photodetector \cite{rodriguez2007nbn,chen2021demonstration,maimon2006nbn,kim2012long,haddadi2017extended,baril2016bulk,craig2013mid}, in which the barrier is sandwiched between two n-type regions, features a distinct design that is less susceptible to crystalline defects and effectively reduces the dark current and noise brought on by the Shockley-Read-Hall (SRH) generation, surface states, and various other processes. Due to the high cost of fabrication and the complexity of such structures, an accurate theoretical modeling is essential for developing the physics of such IR photodetectors \cite{kumar2023advancing}. In this work, we develop an accurate physics-based theoretical reliable simulation model and propose a novel device design for the nBn photodetector \cite{reine2014numerical,maimon2006nbn} that offers relatively high performance and excellent quantum efficiency within the MWIR spectrum \cite{d2012electrooptical}. We then use our in-depth computational analysis to elucidate intriguing physics and predict the various performance limiting factors for an InAs\textsubscript{1-x}Sb\textsubscript{x} (IAS) \cite{krier2007characterization,lackner2009growth,rogalski2020inassb,martyniuk2013SPIE,rogalski1989inas1,shaveisi2023design} based nBn MWIR photodetector, where the barrier is designed with a large band gap Al\textsubscript{0.7}In\textsubscript{0.3}As\textsubscript{0.3}Sb\textsubscript{0.7} (AIAS) \cite{bank2017avalanche,maddox2016broadly,ren2016alinassb} material. \\
\indent The device structure, depicted in Fig.~\ref{preliminaries}, consists of three principal layers: a thick absorber layer (AL) of n-type narrow gap IAS material with a thickness of 2.7 $\mu$m, a barrier layer (BL) of n-type lattice-matched AIAS material with a thickness of 0.25 $\mu$m, and a contact layer (CL) of n-type IAS with a thickness of 0.27 $\mu$m.
%, which in short termed as n (n-type AL) - B (BL) - n (n-type CL) structure.
The thickness of the BL (t\textsubscript{BL}) is considered to be sufficiently large to inhibit electron tunneling between the CL and the AL layers. As a result, the majority current is impeded by the barrier material when a proper bias is applied. The absence of a significant electric field in the narrow gap material prevents the SRH generation and the Band-to-band (BTB) tunneling thereby, the nBn photodetectors operating in the MWIR region exhibit lower levels of dark currents and noise \cite{martyniuk2013SPIE}. This characteristic enables a reduction in the cooling demands associated with these devices. 

\begin{figure*}[!htbp]
	\centering
        \subfigure[]{\includegraphics[height=0.24\textwidth,width=0.30\textwidth]{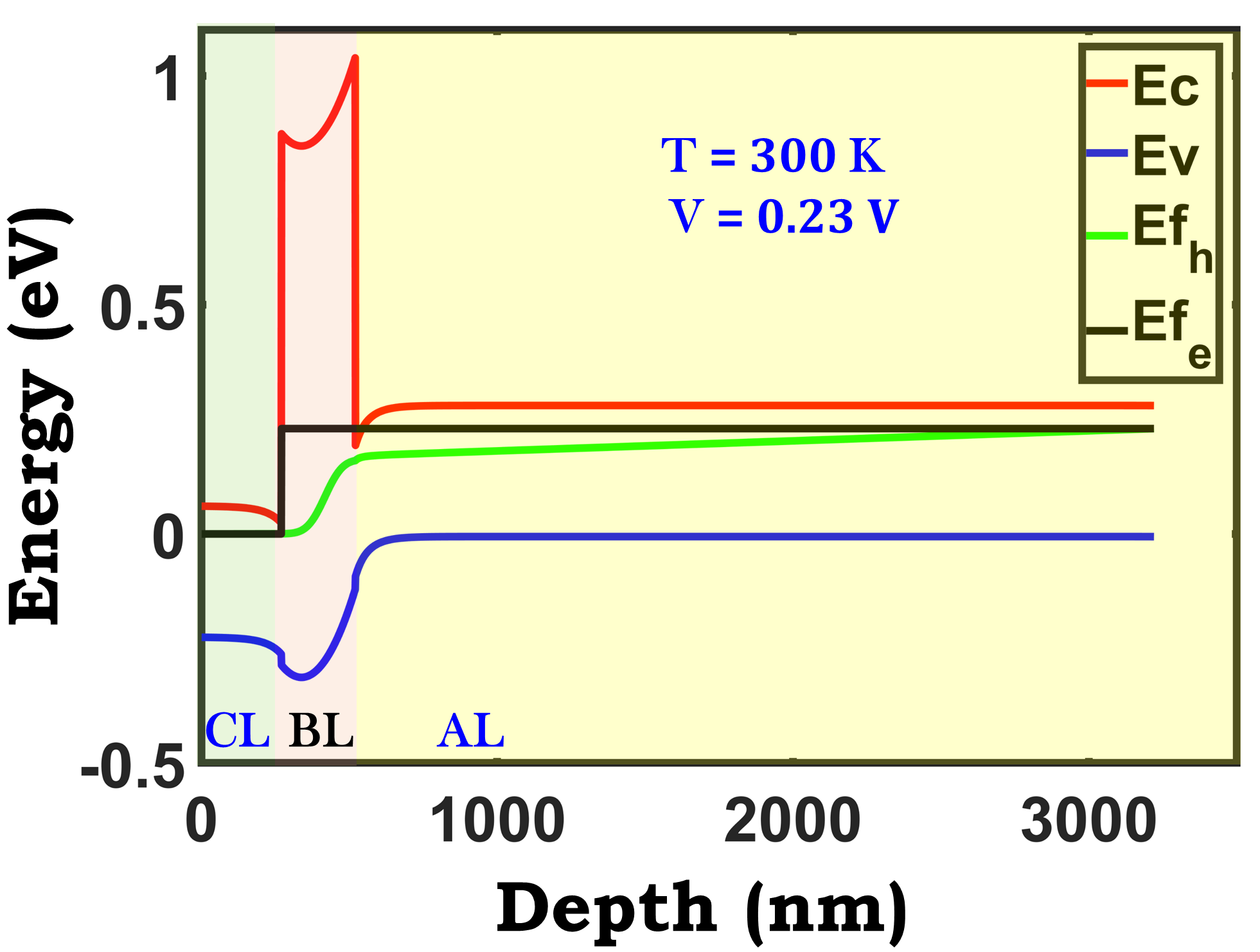}}\label{at positive bias}
	\quad
	\subfigure[]{\includegraphics[height=0.24\textwidth,width=0.30\textwidth]{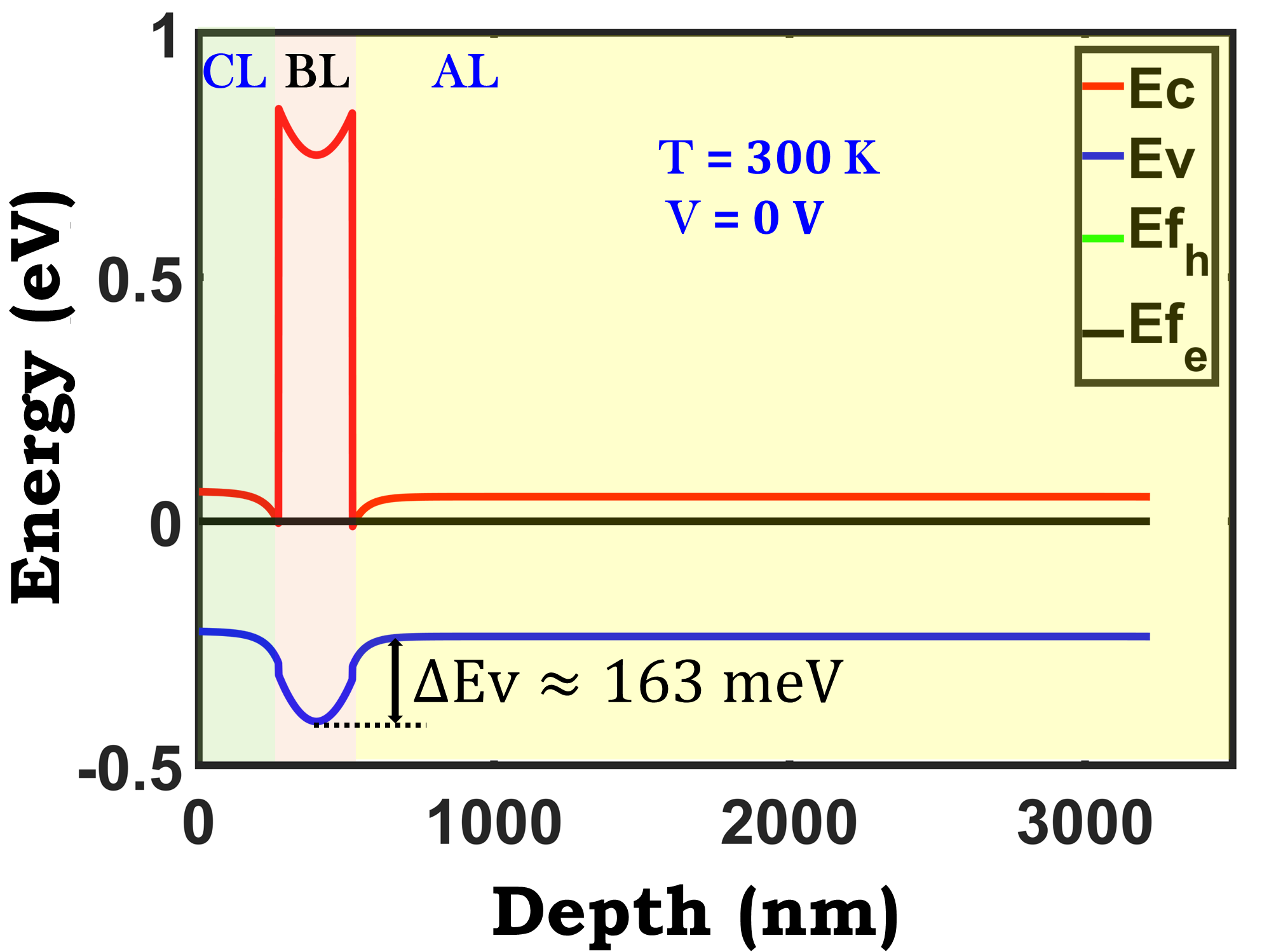}}\label{at zero bias}
	\quad
	\subfigure[]{\includegraphics[height=0.24\textwidth,width=0.30\textwidth]{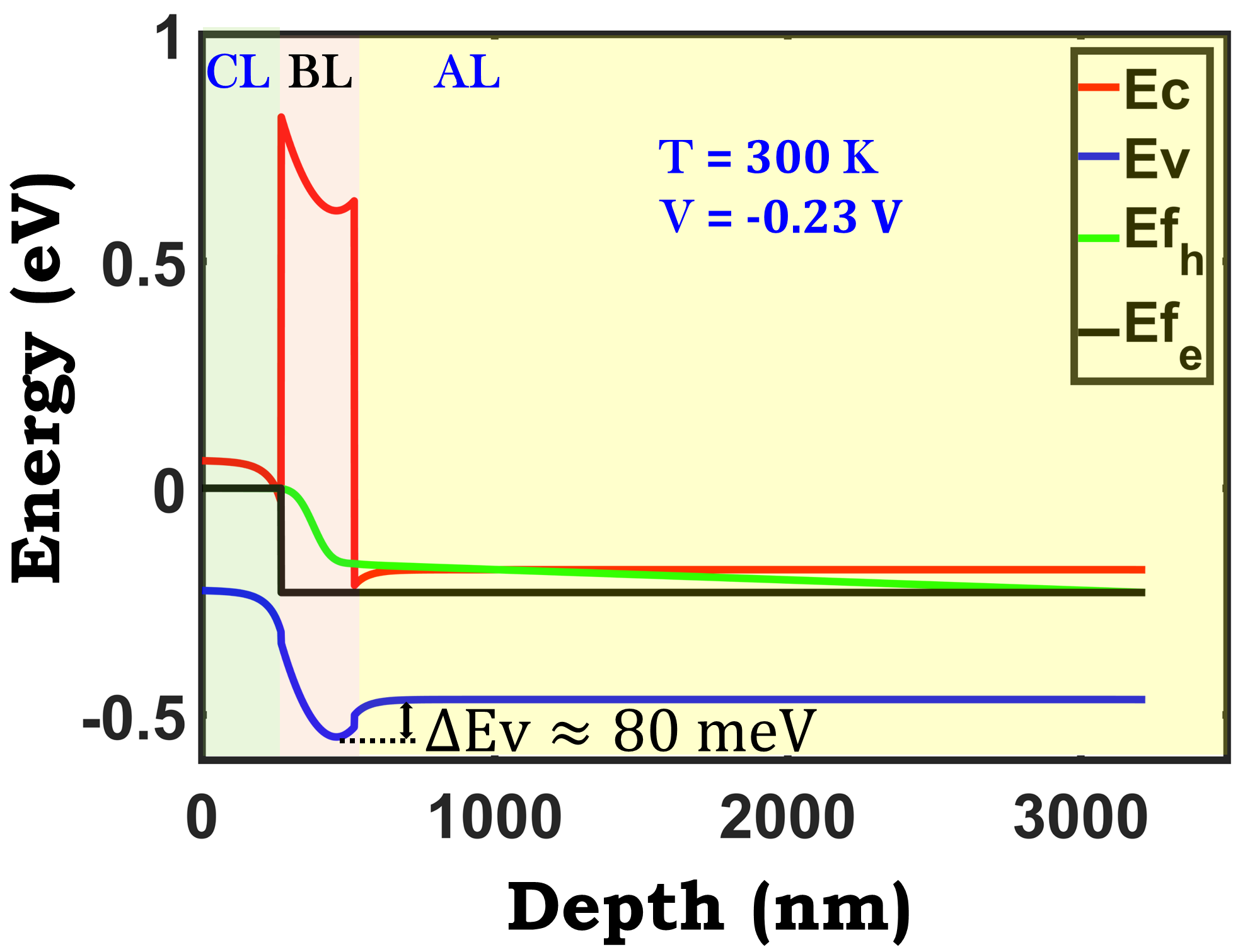}}\label{at negative bias}
	%\quad
	\caption{Energy band profiles for the nBn photodetector under (a) positive bias, (b) equilibrium, i.e., zero bias, and (c) negative applied bias voltage. The high potential barrier in the VB that appears within the BL at zero bias prevents holes from moving to the CL on the left. Under equilibrium, Ef\textsubscript{h} is superimposed to Ef\textsubscript{e} due to their identical spatial alignment. At negative bias, the holes will be able to traverse the potential barrier and reach the CL. The presence of a strong electric field within the barrier region facilitates the movement of carriers, thus contributing to the overall drift current. The electrons within the CL are unable to traverse to the AL due to the large CBO.}
	\label{potential energy profile}
\end{figure*}

\begin{figure}[!t]
	\centering
    \subfigure[]{\includegraphics[height=0.28\textwidth,width=0.45\textwidth]{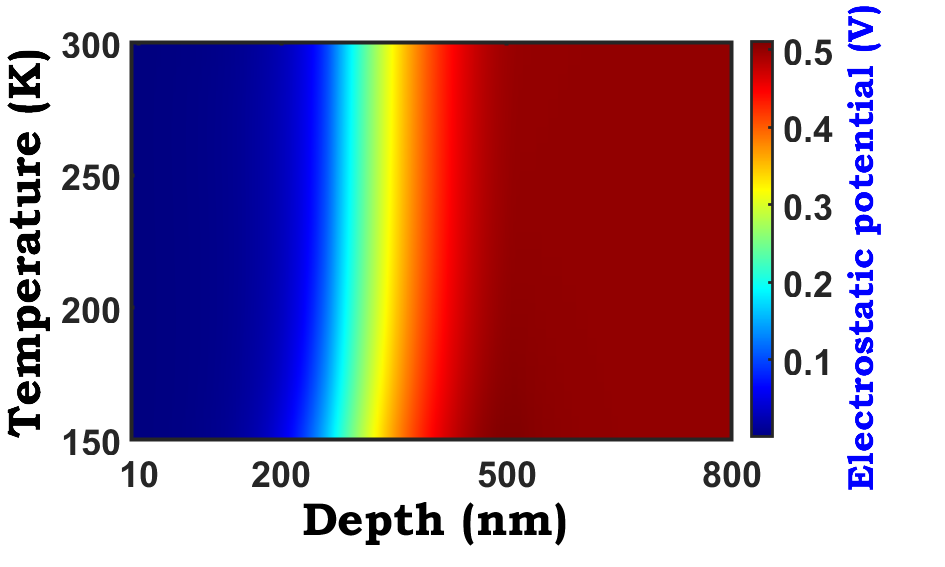}}\label{Electrostatic potential}	
    
  \subfigure[]
{\includegraphics[height=0.30\textwidth,width=0.45\textwidth]{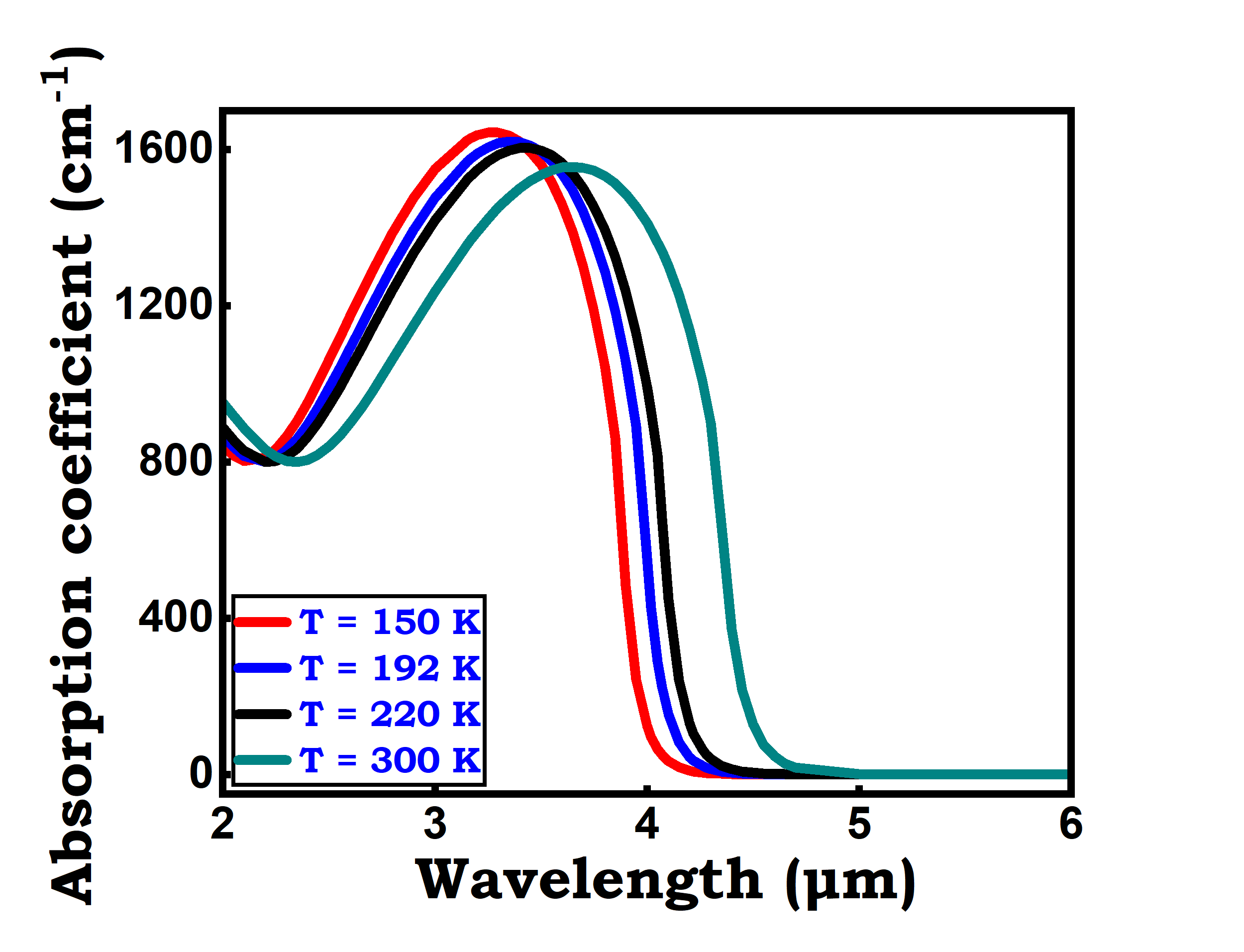}}
	\quad
	\caption{Electrostatic potential and the absorption coefficient of the MWIR nBn photodetector under consideration (a) variation in the electrostatic potential across the device structure as a function of temperature with an applied bias of -0.5 V (b) behavior of the absorption coefficient, $\alpha$, in the absorber region as a function of the incoming radiation wavelength at various temperatures.}
	\label{absorption coefficient}
\end{figure}

\indent Our methodology comprises the finite difference method in conjunction with the linear interpolation technique to solve the Poisson and the continuity equations for the carriers while taking into account the temperature, doping, and structural parameters of the ternary alloy material system. We obtain the electrostatic potential of the heterojunction, hole quasi-Fermi-level outside the thermal equilibrium to build the band structure of the considered device design. The BTB tunneling, the trap-assisted tunneling (TAT), the SRH G-R, and the Auger G-R processes are some of the primary sources of the dark current, as shown in Fig. \ref{preliminaries} (a). \\
\indent Incorporating a BL into the design of IR detectors has the potential to reduce the unfavorable extrinsic SRH G-R contribution substantially. In this study, we thereby focus on the radiative recombination, the Auger G-R, and other thermally generated processes that exhibit dominance in nBn photodetectors. The device layout using GaSb as the substrate is shown in Fig. \ref{preliminaries} (b). The n$^{+}$ CL (i.e., bottom contact) made up of IAS material is the collector layer for the photogenerated electrons, which also serves as the buffer layer to reduce epitaxial strain between the GaSb and the AL for the device. The nBn structure is shown on top of this buffer layer. We use an iterative approach to solve the specific equations and relations used in the device’s modeling. The entire sequence of the numerical simulation is outlined in Fig. A1. The performance of the proposed device is analyzed based on the various calculated transport and optoelectronic parameters such as the carrier density, the electric field, the electrostatic potential, the absorption coefficient for the absorber region, the dark current density, the responsivity, and the quantum efficiency, etc., as a function of the applied bias voltage, operating temperatures, and the structural parameters. The specifications for the CL, BL, and AL to design the considered nBn MWIR photodetectors are given in Tab. \ref{table1}.

\begin{figure*}[!t]
	\centering
	\subfigure[]{\includegraphics[height=0.24\textwidth,width=0.32\textwidth]{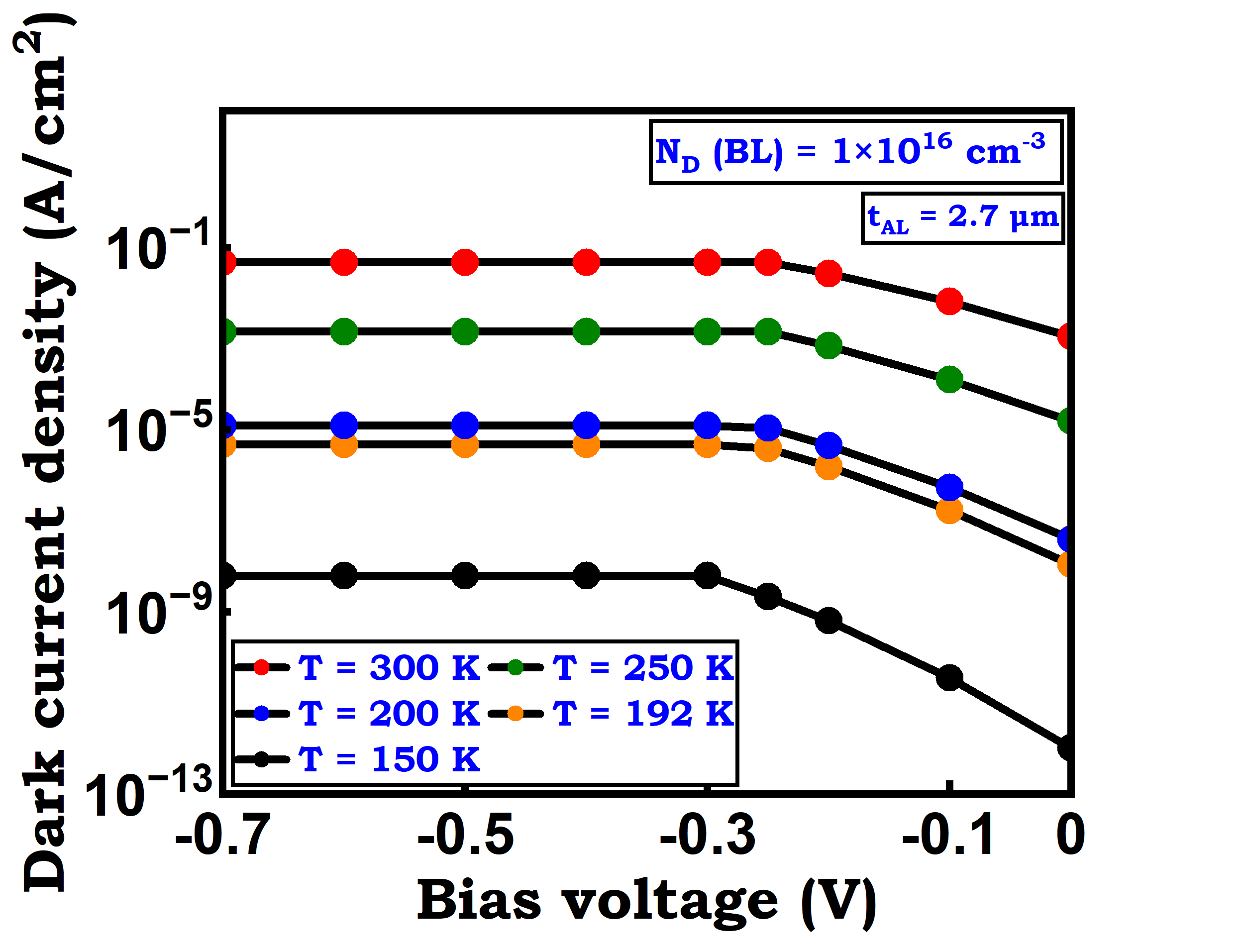}}\label{fixed parameters}
	%\quad
	\subfigure[]{\includegraphics[height=0.24\textwidth,width=0.32\textwidth]{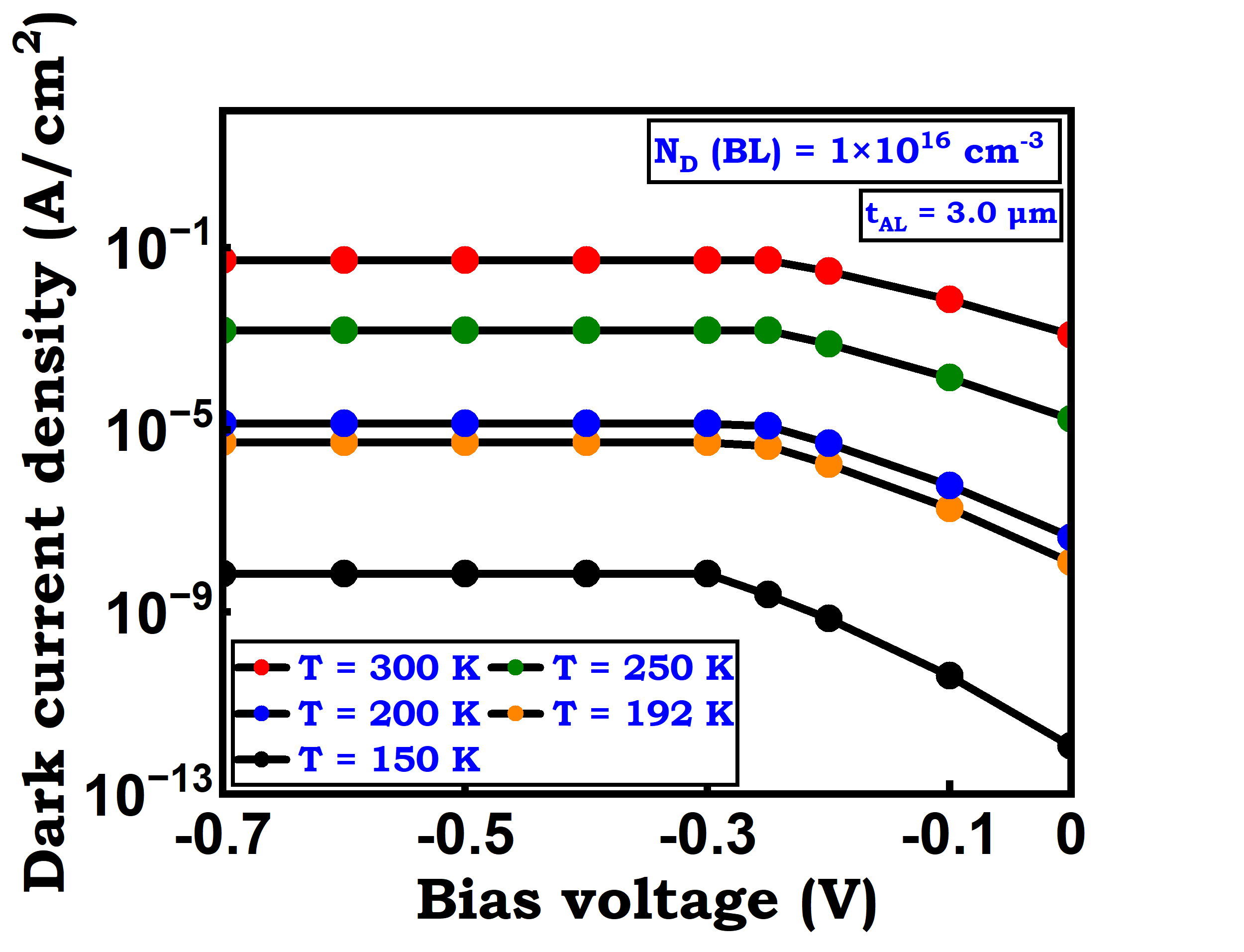}}\label{AL depth effect}
	%\quad
 \subfigure[]{\includegraphics[height=0.24\textwidth,width=0.32\textwidth]{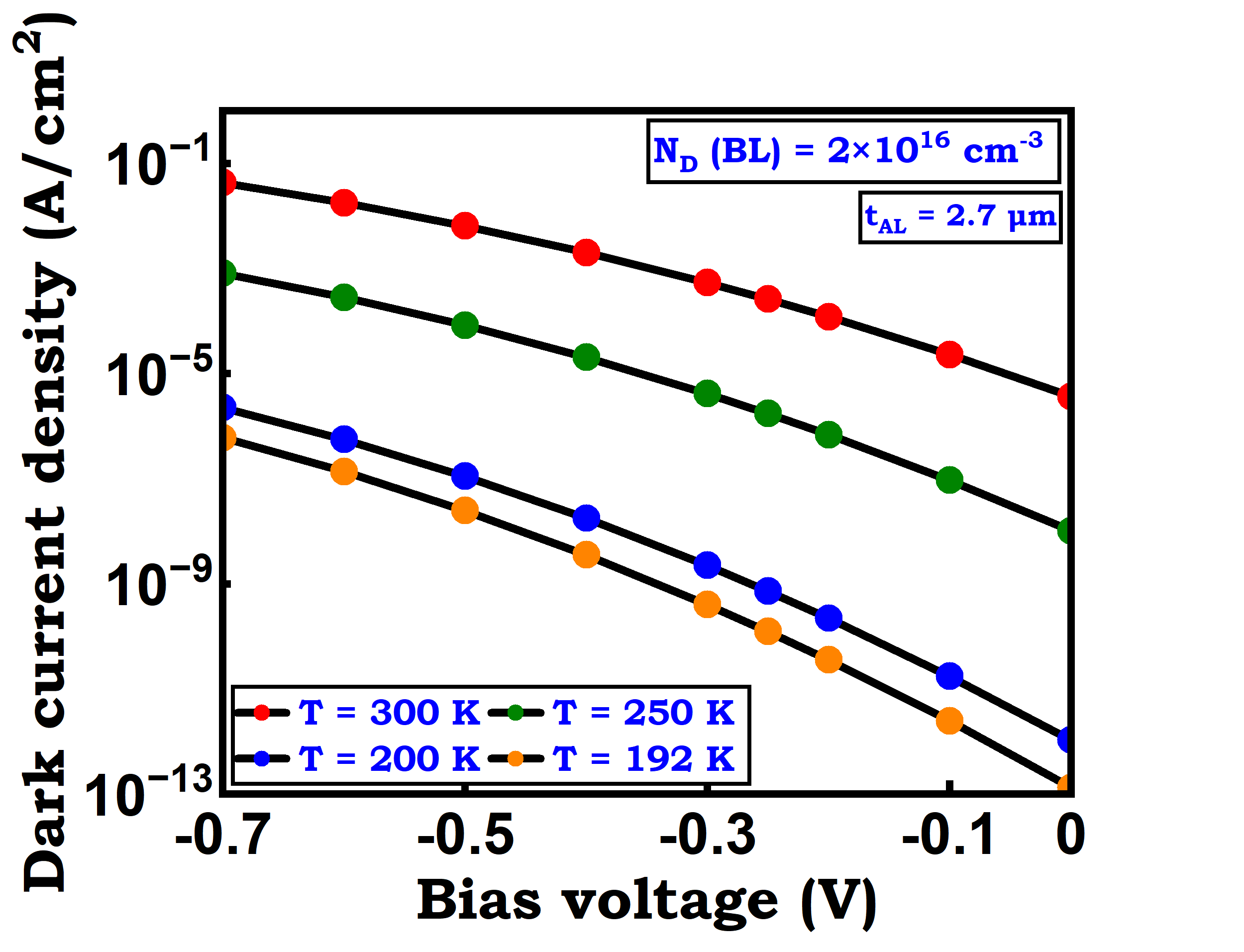}}\label{BL doping effect}
	%\quad
	\caption{Bias dependence of the diffusion-limited dark current density for the nBn photodetectors at various operating temperatures for (a)~$t_{_{AL}} = 2.7~ \mu m $ and $N_{_D} ~(BL)= 1\times 10^{16} ~cm^{-3}$, (b) ~ $t_{_{AL}} = 3.0~ \mu m $ and $N_{_D} ~(BL)= 1\times 10^{16} ~cm^{-3}$, and (c)~ $t_{_{AL}} = 2.7~ \mu m $ and $N_{_D} ~(BL)= 2\times 10^{16} ~cm^{-3}$. The dark current density, as shown in (a) and (b), exhibits an exponential relationship at low bias until it reaches the saturation voltage. Beyond this voltage, the dark current plateaus at a constant value. In contrast, the saturation level shifts to higher voltages when the doping of the BL is increased, as shown in (c).}
	\label{dark current density}
\end{figure*}

Figures \ref{potential energy profile} (a), (b), and (c) depict the potential energy profile of the nBn structure at three distinct applied voltages: positive bias, equilibrium, and negative bias. It is assumed that the electron quasi-Fermi levels in the BL and AL are in equilibrium with each other. As shown in Fig. \ref{potential energy profile} (c), by applying a voltage of V = - 0.23 V to the structure, the energy barrier for holes is lowered by nearly two orders of magnitude (from 163 to 80 meV) compared to an equilibrium condition. It is evident that the applied voltage primarily affects $\Delta E_v$, and when the nBn detector is reverse-biased, it indicates that a positive voltage is applied to the absorber contact. \\
\indent The nBn band structure, as illustrated in Fig. \ref{potential energy profile} (c), hinders the flow of the majority carriers (electrons) through a large conduction band offset (CBO) but enables the flow of the minority carriers (holes) through a near-zero valence band (VB) offset. Therefore, when a relatively low operating voltage is applied, it falls almost entirely across the barrier, thereby separating the photogenerated carriers. 
For $\Delta E_v$ $ > $ 3$k_B$T, the minority carriers in the nBn architecture are effectively blocked therefore, the condition for unhindered minority carrier transport to the CL ($\Delta E_v$ $<$ 3$k_B$T) is met when applied bias exceeds –0.23 V. The band bending phenomena is also observed at the boundaries of the BL, which signifies the accumulation of the majority charge carriers in the vicinity of the space charge region, which spans the depth of the BL. The mobile charges have migrated to the neighboring IAS-based AL and CL, depleting the entire BL. 

\begin{table}[!t]
\caption{\label{table1}
Parameters used to design the device structures. }
\begin{ruledtabular}
\begin{tabular}{ccc}
\bf{$x_{_{Sb}}$} & 0.09 &  \\
\\
\bf{T} & 300 K &   \\
\\
\bf{Cut-off wavelength ($\lambda_c$)} & 4.33 $\mu$m  \\ 
\\ \hline

\hspace{-3.00cm} \bf{Layers} &\hspace{-4.00cm} 
 \bf{Thickness / Depth} & \hspace{-2.00cm} \bf{Doping}    \\ \hline
\\
\hspace{-3.00cm}\bf{CL} & \hspace{-4.00cm} 0.27 $\mu$m & \hspace{-2.00cm} $5\times 10^{15}$ $ cm^{-3}$  \\
\\
\hspace{-3.00cm}\bf{BL} &\hspace{-4.00cm} 0.25 $\mu$m &\hspace{-2.00cm} $8\times 10^{15}$ $ cm^{-3}$  \\
\\
\hspace{-3.00cm}\bf{AL}  &\hspace{-4.00cm} 2.7 $\mu$m &\hspace{-2.00cm} $1\times 10^{16}$ $ cm^{-3}$  \\
\\
\end{tabular}
\end{ruledtabular}
\end{table}

In Appendix \ref{methodology}, an in-depth investigation of the electric field and the carrier density is provided. There, we demonstrate how the electric field and the carrier density in the nBn photodetector are affected by the temperature and applied bias.

In Fig. \ref{absorption coefficient} (a), we show the distribution of the electrostatic potential across the nBn detector as a function of the temperature and the layer thicknesses, which has a direct impact on the carrier concentration. An inference can be made that as the temperature increases, the electrostatic potential becomes more concentrated within the width of the BL. The potential drop in the AL varies hardly with temperature therefore, the complete depth of the AL is not shown, whereas a noticeable change can be seen in the CL. 

\begin{figure*}[!t]
	\centering
    \subfigure[]{\includegraphics[height=0.26\textwidth,width=0.33\textwidth]{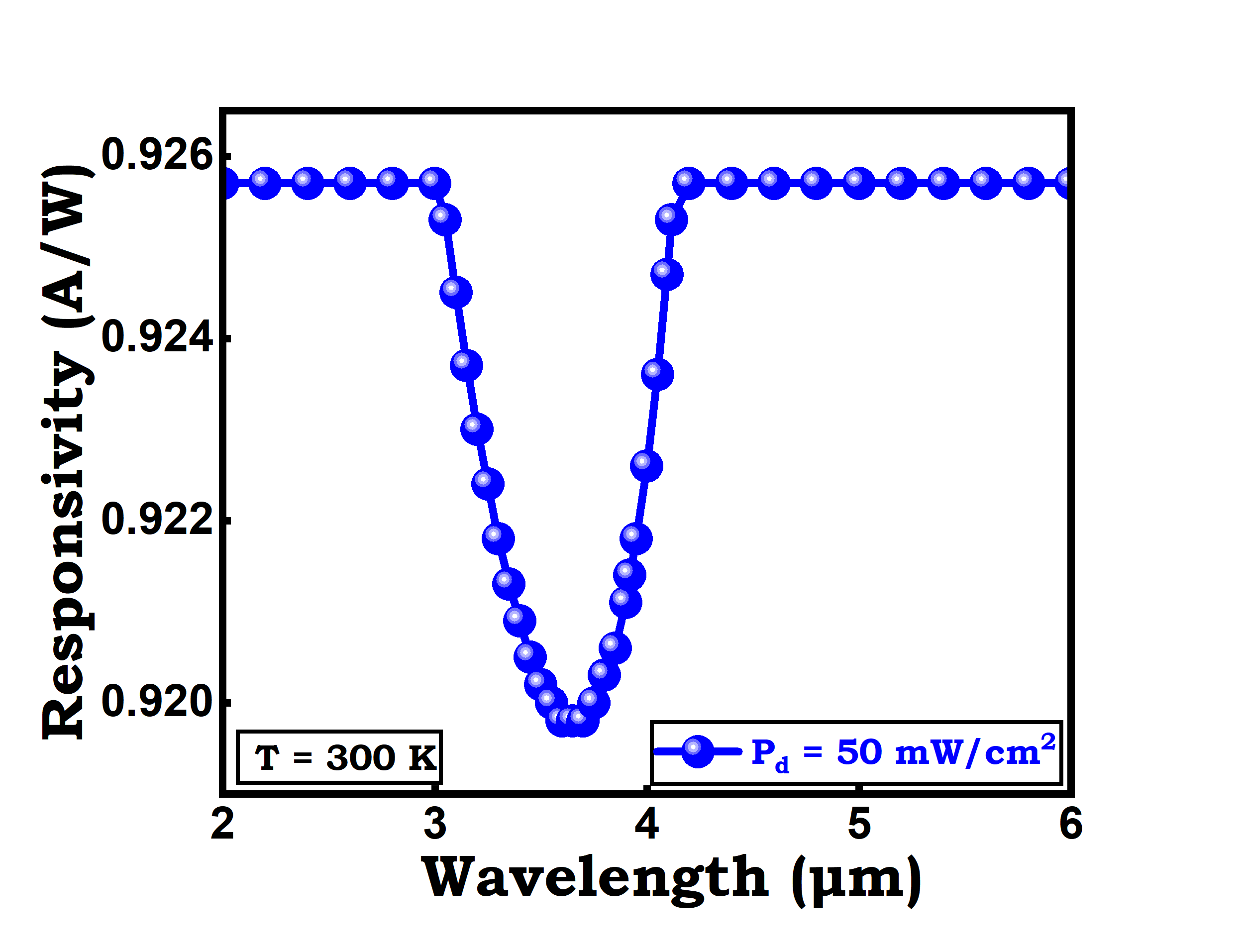}}\label{responsivity plot}
	%\quad
	\subfigure[]{\includegraphics[height=0.26\textwidth,width=0.33\textwidth]{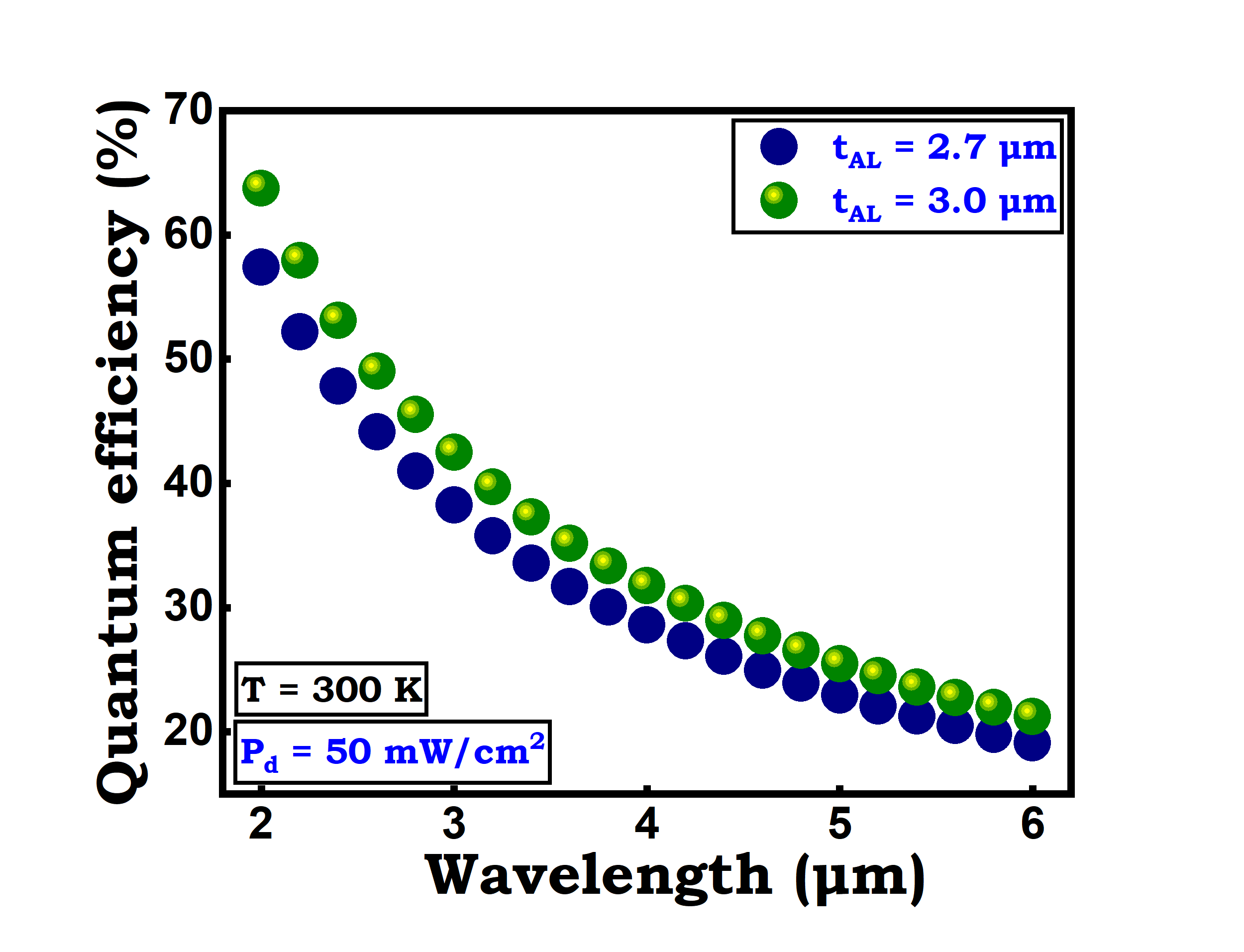}}\label{QE vs lambda}
	%\quad
	\subfigure[]{\includegraphics[height=0.26\textwidth,width=0.32\textwidth]{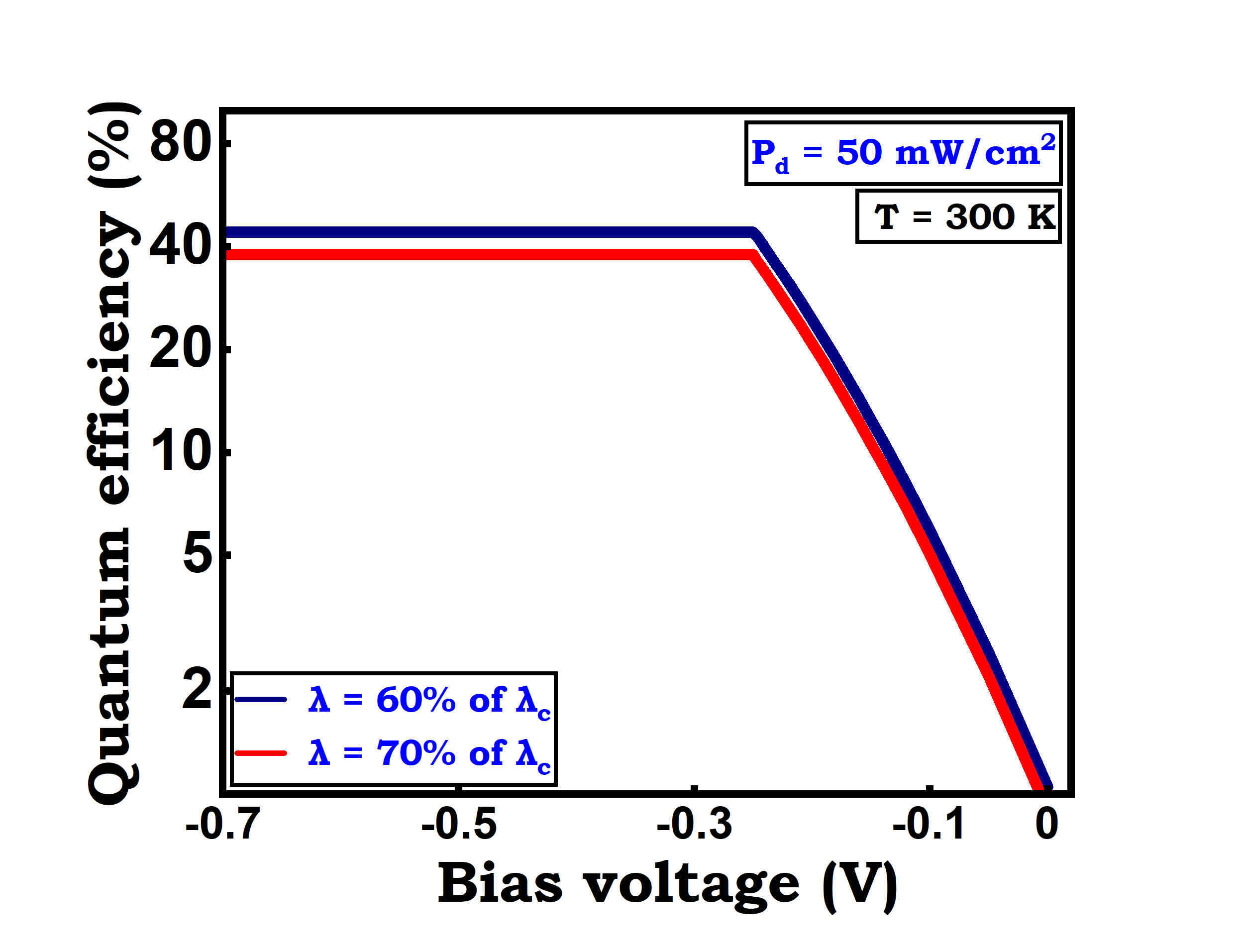}}\label{QE vs bias voltage}
	%\quad
	\caption{ Room temperature responsivity and the quantum efficiency for the MWIR nBn photodetector (a) responsivity as a function of the incoming radiation wavelength (b) wavelength dependence of the quantum efficiency at various AL thicknesses (c) bias dependence of the quantum efficiency at various percentages of cut-off wavelength.}
	\label{RQE}
\end{figure*}

In Fig. \ref{absorption coefficient} (b), we show the absorption coefficient of the IAS absorber in relation to the incoming radiation wavelength and temperature. The energy threshold of absorption is expected to shift as the band gap of AL material varies with temperature. The absorption coefficient exhibits maximum values within the wavelength range of 3.1 to 4.3 $\mu$m for all simulated temperatures. The corresponding $\alpha$ values fall within the range of 1557-1644 $cm^{-1}$. The absorption coefficient exhibits a decreasing trend in its maximum value as temperature increases. This behavior appears to be affected by the band gap of the absorber material.

In Figs. \ref{dark current density} (a), (b), and (c), we demonstrate the dark current density of the nBn MWIR photodetector. It is important to note that the nBn structure operates in a minority carrier manner, hence the hole transport from AL to CL is the primary cause of the dark current. The ``turn-on voltage'' is assumed to be V $\approx$ –(0.23 - 0.25) V. This is the voltage after which the dark current saturates and $\Delta E_v$ falls below 3$k_B$T.
The dark current density is shown in Fig. \ref{dark current density} (b) when the $t_{AL}$ is slightly increased and held constant at 3 $\mu$m. 
Figure \ref{dark current density} (c) demonstrates the dark current density when the BL doping is increased from $1\times 10^{16}~cm^{-3}$ to $2\times 10^{16}~cm^{-3}$. Changing the thickness of the AL has a low impact, while changing the doping of the BL has a noticeable effect, as shown in Figs. \ref{dark current density} (b) and (c), respectively. The dark current density is shown to be highly voltage-dependent between 0 to –0.23 V. Alternatively, it is less sensitive to voltage changes between -0.23 V and -0.7 V, as shown in Figs. \ref{dark current density} (a) and (b). As previously mentioned, the $\Delta E_v$ value for V = -0.23 V is comparable with 3$k_B$T, indicating that the holes are almost freely transported to the CL and hence contribute to the net dark current. It is observed that for reverse voltages V $<$ –0.23 V, the dark current density rises sharply due to a rapid rise in the hole concentration, whereas above V $>$ –0.23 V, the dark current approaches saturation. But when BL doping is increased, the saturation level shifts to the higher voltages, as shown in Fig. \ref{dark current density} (c). Moreover, the diffusion-limited barrier structure prevents tunneling in the simulated voltage range.

The spectral response of the considered nBn photodetector as a function of the wavelength at a temperature of 300 K for a constant bias of -0.25 V with a radiation power density of 50 mW/$cm^{2}$ is presented in Fig. \ref{RQE}(a). The maximum responsivity obtained at lower and higher wavelengths within the MWIR spectrum is recorded as 0.9257 A/W. The same responsivity value is also recorded at $\lambda_c$ = 4.33 $\mu$m. The nBn device here consist $t_{AL}$ of 2.7 $\mu$m. The VB barrier reduces carrier collection, resulting in slightly lower responsivity values close to the cut-off wavelength. The effects of photogenerated carriers altering the nBn barrier height at higher temperatures could be responsible for the slight increase in the responsivity observed at shorter wavelengths. Due to its large responsivity and low dark current density, the proposed nBn design has a very high external quantum efficiency. 

The quantum efficiency, which in turn affects the flux of the photogenerated carriers transported to the CL, is directly affected by the $\Delta E_v$. The dependence of the quantum efficiency on the wavelength and applied bias voltage is depicted in Figs. \ref{RQE} (b) and (c). In Fig. \ref{RQE} (b), we present the dependence of the quantum efficiency on the wavelength at a temperature of 300 K for two different AL thicknesses. The photodetector exhibits a maximum efficiency of 57.39\% and 63.74\% at a bias voltage of -0.25 V for 2.7 $\mu$m and 3.0 $\mu$m AL thicknesses, respectively, with a radiation power density of 50 mW/$cm^{2}$. In Fig. \ref{RQE} (c), we show the quantum efficiency at two different wavelengths, 60\%, and 70\% of the $\lambda_c$ and it rises sharply to a maximum value of 44.18\% and 37.87\%, respectively, at voltage nearly -0.25 V as the reverse bias increases. Beyond this threshold of applied reverse bias, the quantum efficiency remains unaffected by the VB offset. Since there is no tunneling contribution and the dark current density slightly saturates, showing the photoconductive effect, while quantum efficiency reaches its maximum value, it can be stressed that such nBn structure may be operated above -0.25 V. \\
\indent In conclusion, a physics-based theoretical simulation model has been developed via an iterative approach for the Poisson and the continuity solver to develop a framework for the nBn MWIR photodetectors. A remarkable maximum efficiency of 57.39\% was achieved at room temperature when applying a bias of -0.25 V, coupled with a radiation power density of 50 mW/$cm^{2}$. We have recorded the maximum quantum efficiency of 44.18\% and 37.87\% at 60\% and 70\% of the $\lambda_c$, respectively. Furthermore, a maximum responsivity of 0.9257 A/W has been recorded within the MWIR spectrum. Through a comprehensive analysis, we have demonstrated that our proposed device design effectively reduces the dark current density by confining the electric field inside the barrier while preserving a superior level of quantum efficiency, and the current in such detectors is diffusion-limited. Hence, the G-R and tunneling currents do not limit the high performance of the nBn architecture. Insights uncovered here could be of broad interest to critically evaluate the potential of the nBn structures for MWIR IR photodetectors.
%\section*{Supplementary material}
%See the supplementary material for the complete details about various calculations, relevant equations, and simulation methodology.\\
\section*{Acknowledgments} 
The authors acknowledge funding from ISRO, under the ISRO-IIT Bombay Space Technology Cell. B.M. also acknowledges funding from the Science and Engineering Research Board (SERB), Government of India, under Grant No. CRG/2021/003102.

\section*{Author declarations}
\subsection*{Conflict of Interest}
The authors have no conflicts to disclose.
\subsection*{Author Contributions}
\noindent {\bf{Rohit Kumar:}} Investigation (equal); Conceptualization (equal); Visualization (lead); Data curation (lead); Methodology (lead); Formal analysis (lead); Validation (lead); Writing-original draft (lead); Writing-review \& editing (lead). {\bf{Bhaskaran Muralidharan:}} Investigation (equal); Conceptualization (equal); Funding acquisition (lead); Project administration (lead); Resources (lead); Supervision (lead); Writing- review \& editing (equal). 

\section*{Data Availability}
The data that support the findings of this study are available from the corresponding author upon reasonable request.\\

\appendix
\section{Methodology}\label{methodology}
\setcounter{figure}{0} 
\setcounter{table}{0}
\renewcommand{\figurename}{FIG. A\hskip-\the\fontdimen2\font\space}

We solve the Poisson and the continuity equations for the carriers using the finite difference method and obtain the electrostatic potential of the heterojunction, hole quasi Fermi-level outside thermal equilibrium to build the band structure of the considered device design. We assume that no G-R processes are taking place within the BL, and the absorber region is uniformly illuminated. The entire sequence of the numerical simulation is outlined in Fig. A1. The subsequent equations presented herein are relevant for the calculation of various transport and optoelectronics parameters associated with the MWIR nBn photodetectors.

\begin{figure}[!t]
	\centering
	{\includegraphics[height=0.85\textwidth,width=0.50\textwidth]{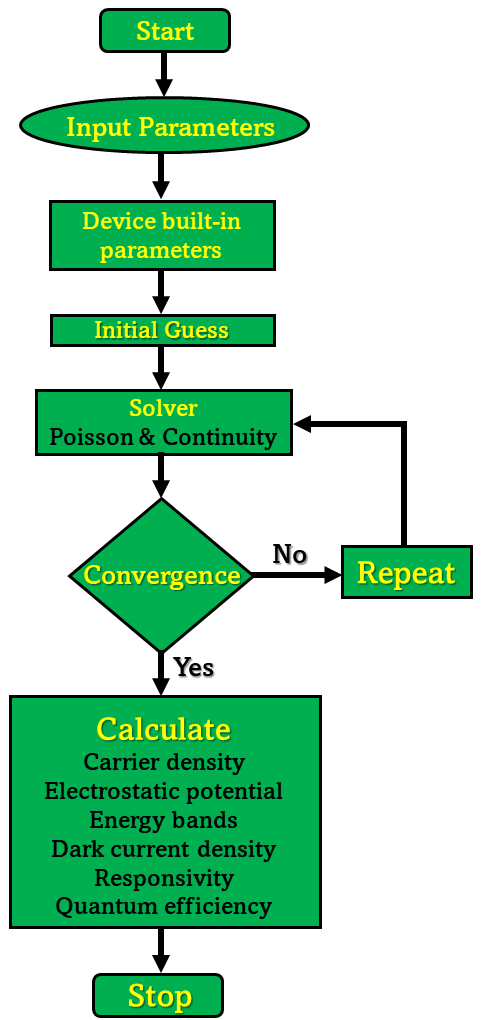}}
	\quad
	\caption{Flowchart to calculate the transport and optoelectronics parameters using the Poisson and continuity solver.}
	\label{Flowchart}
\end{figure}

By using the linear interpolation technique, the lattice constant or the mobility of IAS can be expressed as \cite{adachi2017iii}
\begin{equation}
    (a/\mu)_{_{IAS}}=x_{_{Sb}}\times (a/\mu)_{_{InSb}} + (1-x_{_{Sb}})\times (a/\mu)_{_{InAs}}\:,
	\label{lattice constant_IAS}	
\end{equation}

\noindent where $a$ denotes the lattice constant, $\mu$ represents the mobility of the carriers and $x_{_{Sb}}$ is the molar fraction. In order to determine the mobility at any given temperature, the mathematical expression is employed as \cite{schuster2014analysis}

\begin{equation}
    \mu|_{_T}=\mu|_{_{(300~K)}} \times \Big[\dfrac{T}{300}\Big]^{-\zeta}\:,
	\label{mobility at various T}	
\end{equation}

\noindent where $\mu|_{_{(300~K)}}$ is the mobility of the carriers at room temperature. The value of $\zeta$ can be extracted from TABLE S\ref{table1}. The diffusion current in the nBn photodetector is limited by the intrinsic carrier concentration $n_i$ of the IAS material, which can be calculated as\cite{rogalski1989intrinsic,rogalski1989inas1}

\begin{equation}
\begin{aligned}
   (n_i)_{_{IAS}}=(8.50~x_{_{Sb}} - 6.73~x_{_{Sb}}^2 -1.53\times10^{-3}x_{_{Sb}}~T\\
   + 4.22\times10^{-3}~T + 1.35)E_g^{0.75}~T^{1.5}\exp \Big(\dfrac{-E_g}{2k_BT}\Big)\times 10^{14}\:,
   \end{aligned}
	\label{intrinsic carrier conc.}	
\end{equation}

\noindent where $E_g$ is the bandgap of the material. The band gap of the IAS-based CL and the AL as a function of molar composition and temperature has been calculated using the following relation \cite{wieder1973photo,shaveisi2023design,schuster2014analysis,rogalski1989inas1,d2012electrooptical}

\renewcommand{\tablename}{TABLE A\hskip-\the\fontdimen2\font\space}
\begin{table*}[!t]
\caption{\label{table1_append}
Parameters for the binary materials at T = 300 K \cite{schuster2014analysis,rogalski1989inas1,adachi2017iii,InSb,InAs,vurgaftman2001band,sai1977new,rogalski2020inassb,martyniuk2014new}.}
\begin{ruledtabular}
\begin{tabular}{cccccc}

\bf{Material}& \bf{$a$ (\AA) }  & \bf{$\mu_e~( cm^2~V^{-1}~s^{-1}$)} & \bf{$\mu_h~( cm^2~V^{-1}~s^{-1}$)} & \bf{$\zeta_e$}  & \bf{$\zeta_h$} \\ 
\hline
\\
\bf{InSb} & 6.4794 & $ 8\times10^{4}$ & 800 & 1.8572 & 1.8572 \\
\\
\bf{InAs} & 6.0583 & $ 3\times10^{4}$ & 500 & 0.7212 & 0.5097 \\
\\

\end{tabular}
\end{ruledtabular}
 \end{table*}

\renewcommand{\tablename}{TABLE A\hskip-\the\fontdimen2\font\space}
\begin{table*}[!t]
\caption{\label{table2_append}
Parameters used for the device simulations \cite{lackner2009growth,IAS,ren2016alinassb,maddox2016broadly,krier2007characterization,bank2017avalanche}. }
\begin{ruledtabular}
\begin{tabular}{ccccc}

\bf{Parameters} & \bf{Unit} & \bf{CL} & \bf{BL}  & \bf{AL} \\ 
\hline
\\
\bf{$E_g|~_{(300~K)}$} & meV & 286 & 1160 & 286 \\
\\
\bf{$\epsilon_r$} & - & 15.298 & 15.5 & 15.298 \\
\\
\bf{$m^*_e$} & $m_0$ & 0.0197 & 0.071 & 0.0197 \\
\\
\bf{$m^*_h$} & $m_0$ & 0.416 & 0.35 & 0.416 \\
\\
\bf{$\mu_h|~_{(300~K)}$} & $cm^2~V^{-1}~s^{-1}$ & $5.27\times 10^2$ & $3\times 10^3$ & $5.27\times 10^2$ \\
\\

\bf{$D_h|~_{(300~K)}$} & $cm^2~s^{-1}$ & 13.636 & 77.625 & 13.636 \\
\\
\hline
\\
\bf{$P_d$} & $mW/cm^{2}$   & \hspace*{-4.0cm} 50  \\
\\
\bf{$|F_1F_2|$} & -  & \hspace*{-4.0cm} 0.2  \\
\\

\bf{$m_0$} & Kg  & \hspace*{-4.0cm} $9.109\times10^{-31}$ \\
\\

\end{tabular}
\end{ruledtabular}
 \end{table*}

\begin{equation}
\begin{split}
    E_g(x_{_{Sb}},T) = 0.411-\bigg[\frac{(3.4\times10^{-4})~ T^2}{(T+210)}\bigg]-0.876\: \\
    x_{_{Sb}} +\: 0.70\: x^2_{_{Sb}} +\: (3.4\times 10^{-4})\: x_{_{Sb}}~T~ (1-x_{_{Sb}})\:,
    \end{split}
	\label{bandgap_IAS}	
\end{equation}
\noindent where $x_{_{Sb}}$ is the molar fraction, $E_g$ is the bandgap, and T is the operating temperature. The absorption coefficient $\alpha$ for the IAS-based AL depends on the incoming radiation wavelength (or photon energy) and temperature and can be calculated using the following equations \cite{d2012electrooptical,shaveisi2023design,schuster2014analysis}.  
When ($E_g$ $\geq$ $h\nu$), the absorption coefficient can be expressed using the Urbach tail model as

\begin{equation}
   \alpha = 948.23~\times~ \exp\:\: [170 ~(h\nu - E_g - 0.001)] \: ,
  \label{absorption coefficent1}	
\end{equation}

\noindent where $h\nu$ is the photon energy. Similarly, when ($E_g$ < $h\nu$), the absorption coefficient can be expressed as \cite{d2012electrooptical,shaveisi2023design,schuster2014analysis} 
\begin{equation}
\begin{split}
   \alpha = 800 ~ + ~ &\frac{K(h\nu - E_g -\Xi)\sqrt{(h\nu - E_g -\Xi)^2 - \Xi^2}}{h\nu} \:,\\
   &\Xi= \frac{E_g}{2} + 0.1 \:,\\
   & K = 10^4~ [1 + 2\: E_g]  \:.
   \end{split}
	\label{absorption coefficent2}	
\end{equation}

We demonstrate the temperature and bias dependence of the carrier density of the nBn photodetector in Fig. A2. The exponential relationship between the carrier concentration and applied bias voltage results in a significant change of the carrier density when there is a relatively small change in the electrostatic potential distribution, as depicted in Fig. A2. 
As a result of the reverse bias applied over the heterojunction, changes are observed in the BL and the accumulation regions. The potential drop is anticipated to occur in the BL because it is the constituent layer with more dopant atoms. An increase in the reverse bias voltage results in a notable increase in the charge density within the accumulation region at the contact-barrier junction. In contrast, a substantial drop in the charge density within the accumulation region at the barrier-absorber junction is observed, resulting in a decrease in the carrier density at this junction. A positive carrier density indicates the presence of ionized donors or holes in that region, whereas a negative carrier density indicates that the majority charge carriers occupy that region. Up to V $\approx$ -0.5 V, the carrier density in the BL is constant, implying that only ionized donors are present. This behavior persists up to a voltage of -0.5 V when the flat band condition is met, indicating that there is no charge density at the barrier absorber edge. Figure A2 (b) provides a clear illustration of the carrier density both before and after it meets the flat band condition. 

We demonstrate the dependence of the electric field on temperature and bias in Fig. A3. The analysis of the nBn photodetector involves considering the electric field, which provides valuable insights into the electrostatic properties. In addition, it is an important consideration in the carrier collection, as a high electric field in the BL may improve the photodetector's response time. The occurrence of undesired phenomena, such as BTB tunneling, can be attributed to the presence of a high electric field. However, this limitation in the nBn photodetector is effectively reduced through the incorporation of a substantial CBO.
The magnitude of the electric field in the AL decreases as the applied bias becomes more negative, as illustrated in Fig. A3. At approximately -0.5 V, the flat band condition takes place, and the electric field is close to zero. As the AL is depleted in response to an increasing negative applied bias, a negative electric field is produced. However, the magnitude of the electric field at the contact barrier junction grows as the applied bias is made more negative.

\begin{figure*}[!t]
	\centering
	\subfigure[]{\includegraphics[height=0.24\textwidth,width=0.32\textwidth]{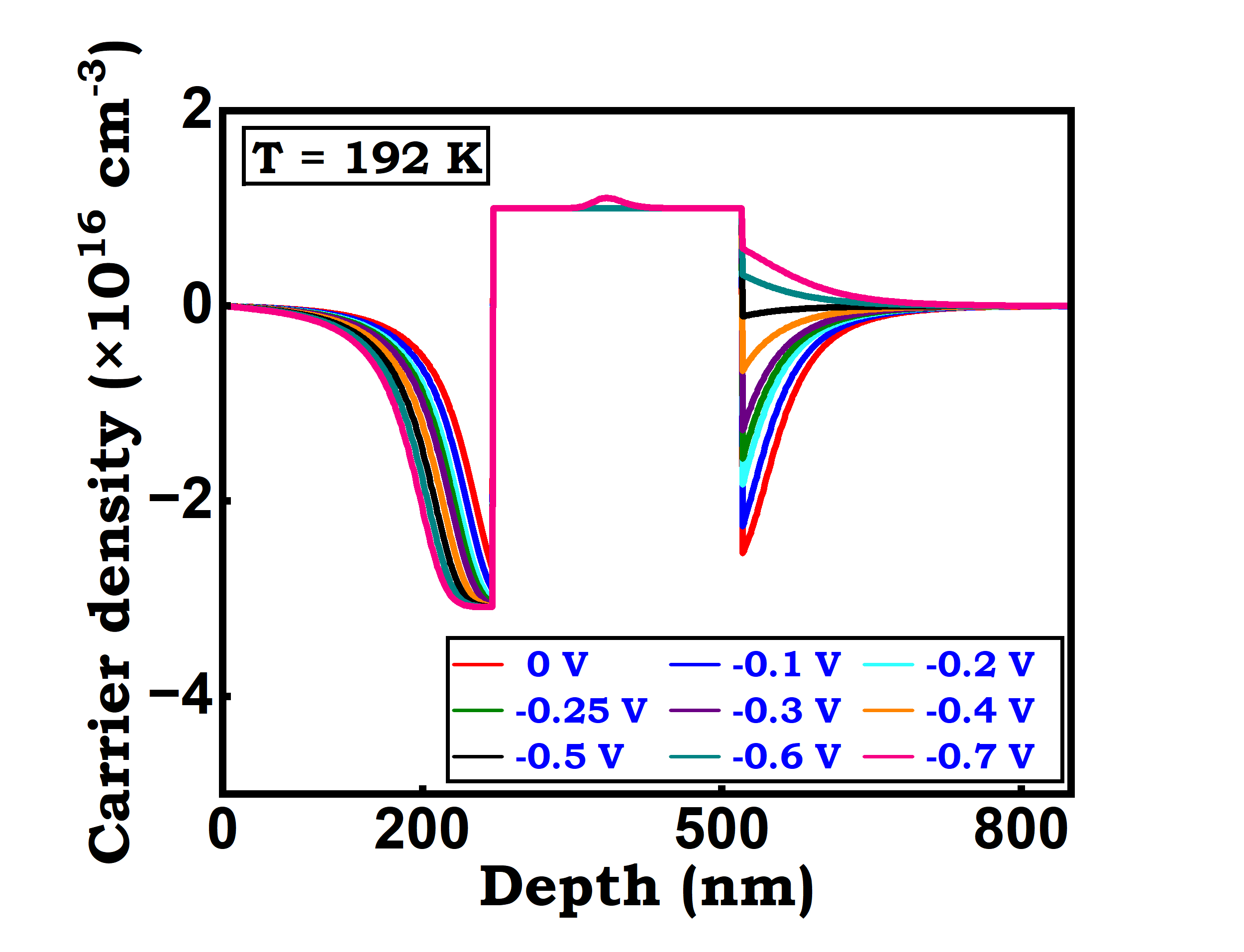}}\label{Carrier density at 192 K}
	%\quad
	\subfigure[]{\includegraphics[height=0.24\textwidth,width=0.32\textwidth]{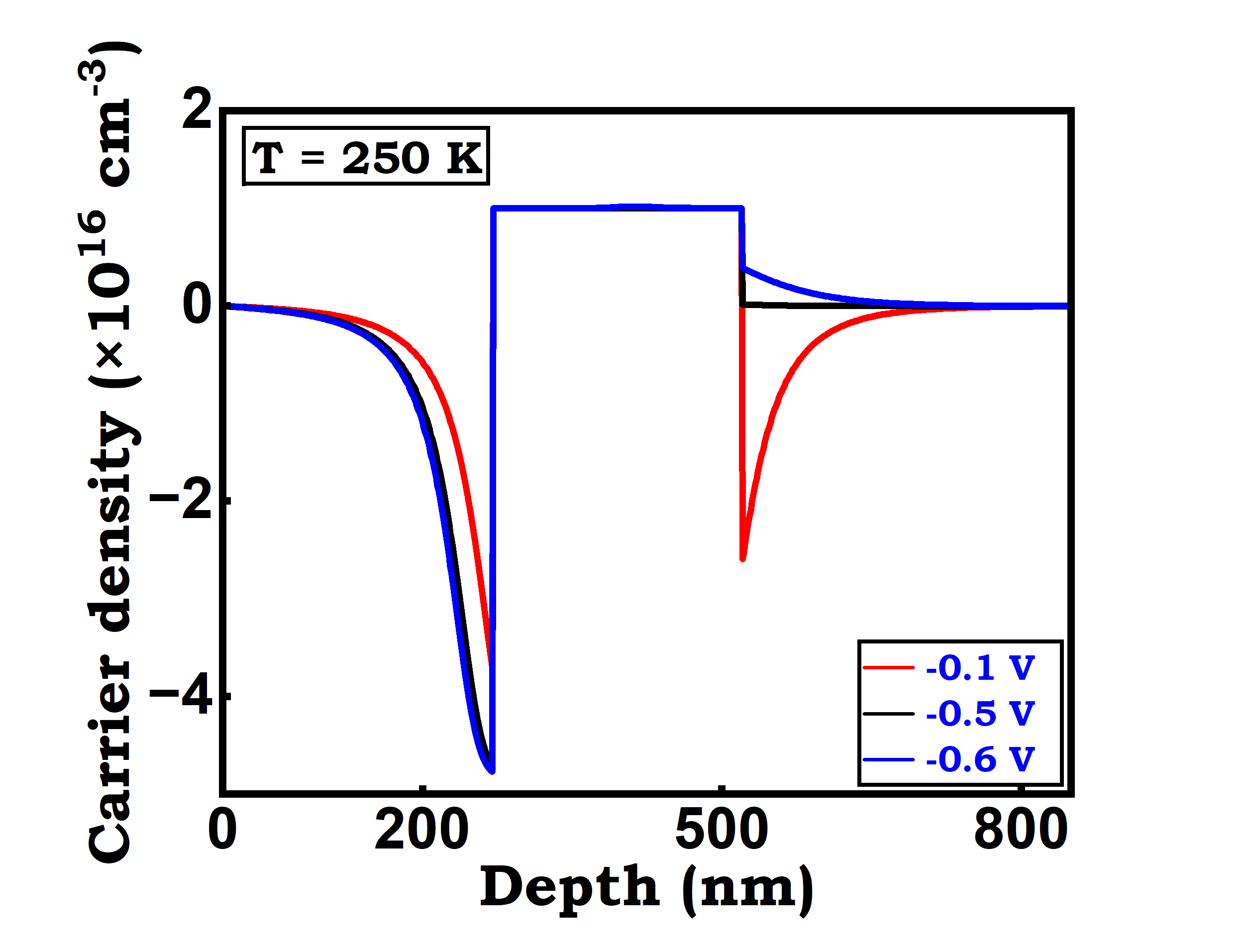}}\label{Carrier density at 250 K}
	%\quad
 \subfigure[]{\includegraphics[height=0.24\textwidth,width=0.32\textwidth]{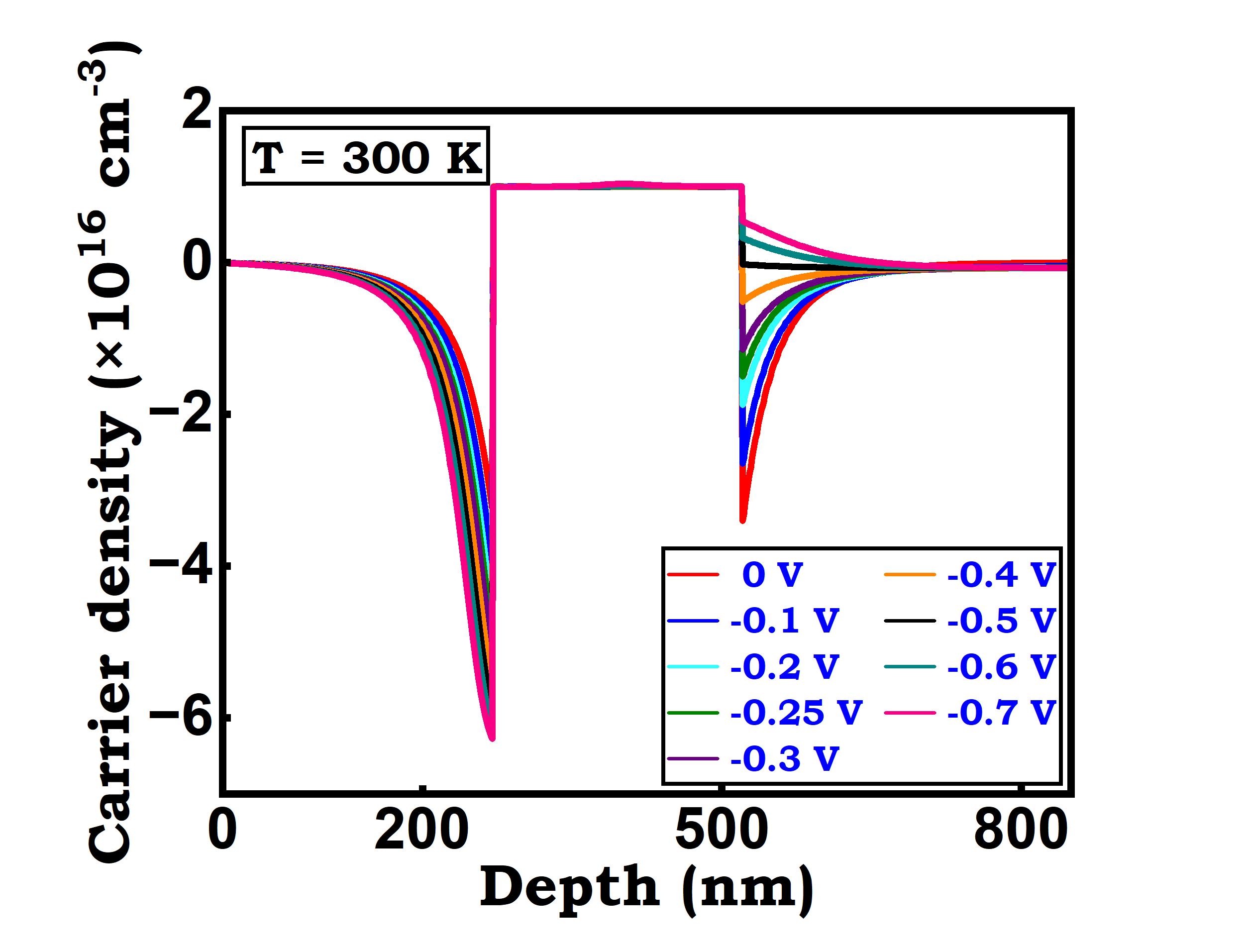}}\label{Carrier density at 300 K}
	%\quad
	\caption{Variations in the carrier density across the heterostructure in relation to the applied bias voltages at (a) T = 192 K, (b) T = 250 K, and (c) T = 300 K. The full extent of the AL region is not shown. Carrier density with a negative value signifies the allocation of electrons in that region, whereas a positive value indicates the presence of holes and positively ionized donors.}
	\label{carrier density}
\end{figure*}

\begin{figure*}[!htbp]
	\centering
	\subfigure[]{\includegraphics[height=0.24\textwidth,width=0.32\textwidth]{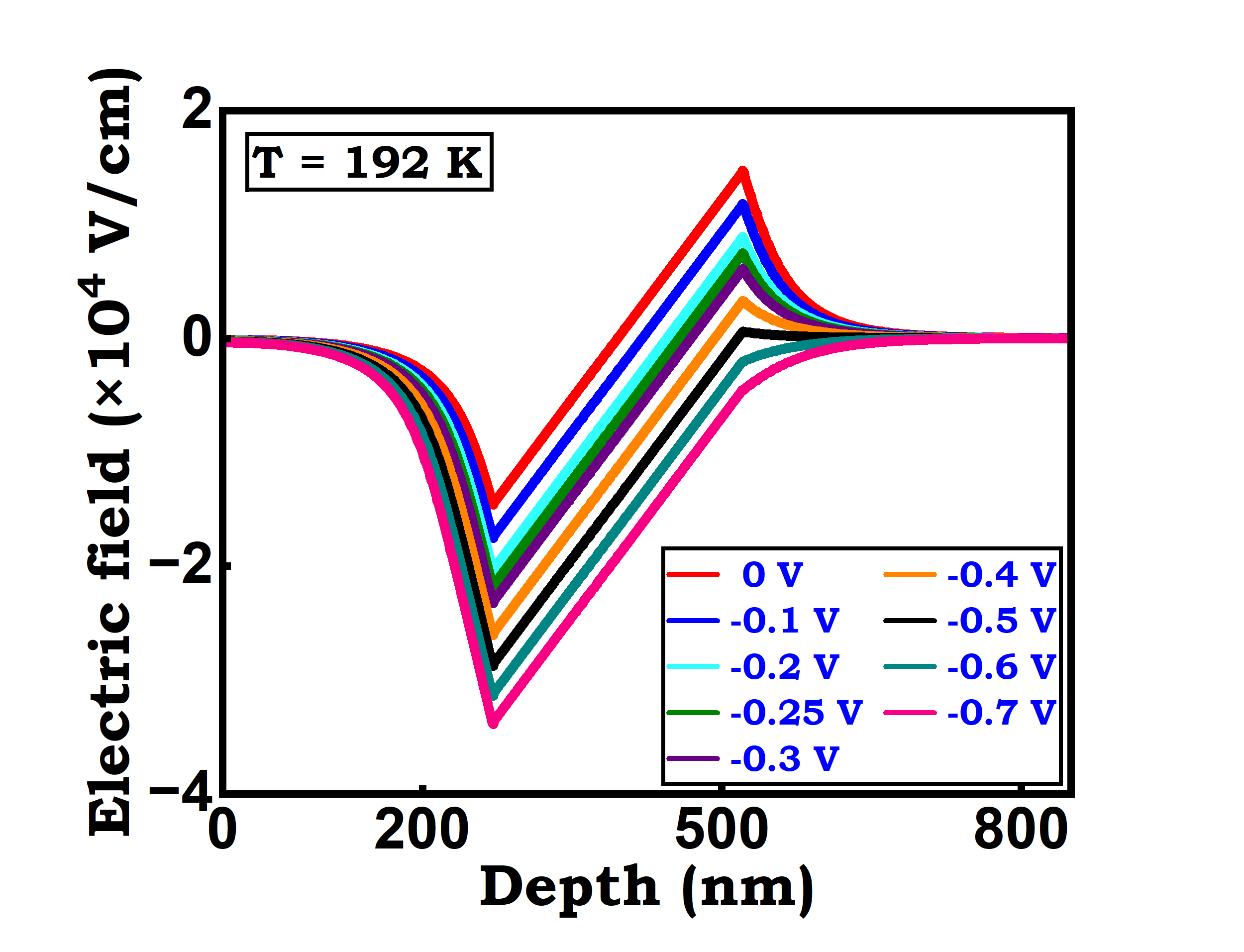}}\label{E-field at 192 K}
	%\quad
	\subfigure[]{\includegraphics[height=0.24\textwidth,width=0.32\textwidth]{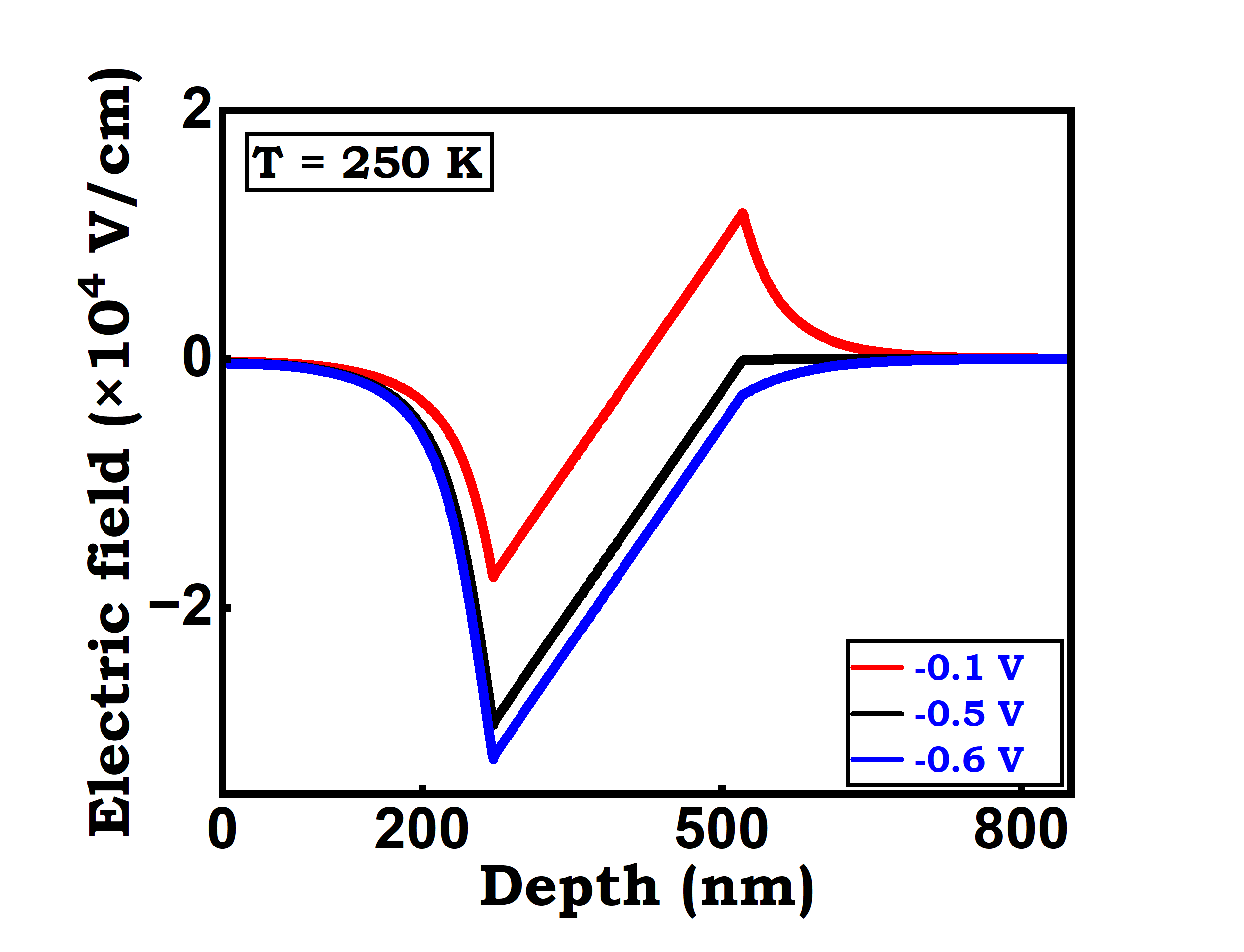}}\label{E-field at 250 K}
	%\quad
 \subfigure[]{\includegraphics[height=0.24\textwidth,width=0.32\textwidth]{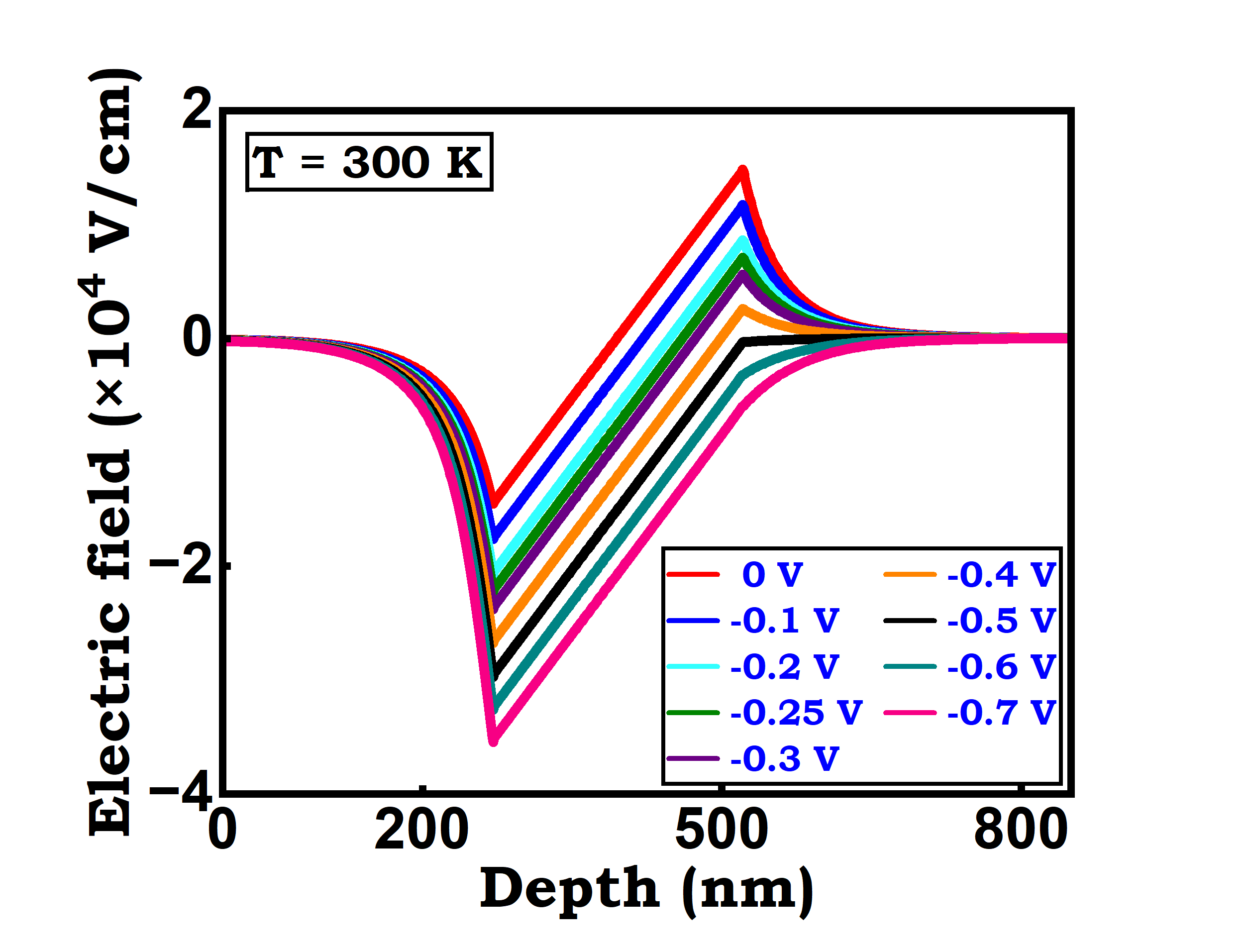}}\label{E-field at 300 K}
	%\quad
	\caption{Bias and position dependence of the electric field across the heterostructure at (a) T = 192 K, (b) T = 250 K, and (c) T = 300 K. The complete width of the AL region is not depicted in the illustration as the electric field remains constant beyond the depletion region.}
	\label{electric field profile}
\end{figure*}

\section{Recombination rate calculations}\label{recombination rate cal}

The radiative recombination mechanism is characterized by the recombination of an electron in the conduction band with a hole in the valence band, resulting in the emission of a photon due to an excess of energy. Therefore, the BTB recombination coefficient, B included in our model, can be written as \cite{bellotti2006numerical,pierret1987advanced,rogalski1989inas1,hall1959recombination} 

\begin{equation}
\begin{split}
B=\Bigg[(5.8  &\times10^{-13})\:\epsilon^{0.5}_\infty \Big(\dfrac{m_0}{m^*_e+m^*_h}\Big)^{1.5}\Big(1+\dfrac{m_0}{m^*_e} +\dfrac{m_0}{m^*_h} \Big)\\
\times &\Big( \dfrac{300}{T}\Big)^{1.5}(E^2_g+3k_B T E_g +3.75~ k^2_B T^2)\Bigg]\:,\\
   & R_{rad} = B\:(np-n^2_i) \:,
         \label{radiative recombiantion rate}	
         \end{split}
\end{equation}

\noindent where $R_{rad}$ is the BTB recombination rate, $k_B$ is the Boltzmann constant, $\epsilon_\infty$ is the high-frequency dielectric constant, and $m^*_e$ and $m^*_h$ are the effective masses of the electrons and holes, respectively. 
Depending on the shape of the bands involved, there are various types of Auger recombination processes. In the context of n-type material, such as the IAS layers studied in the present work, the predominant recombination mechanism is Auger 1. The Auger 1 process, characterized by its non-radiative nature, exhibits dominance at higher temperatures. The Auger carrier coefficients can be defined as \cite{schuster2014analysis} 
%\begin{widetext}
\begin{equation}
\begin{split}
C_n = &\dfrac{\Big(\dfrac{m^*_e}{m_0}\Big)~|F_1 F_2|^2}{2 (n_i ~\epsilon_\infty)^2 ~(3.8\times10^{-18}) \Big(1+\dfrac{m^*_e}{m^*_h}\Big)^{0.5} \Big(1+ 2~ \dfrac{m^*_e}{m^*_h}\Big)}\times \\
&\Big(\dfrac{E_g}{k_BT} \Big)^{-1.5} \exp \Bigg[-\dfrac{\Big(1+2~ \dfrac{m^*_e}{m^*_h}\Big) E_g}{\Big(1+ ~ \dfrac{m^*_e}{m^*_h}\Big)k_B T} \Bigg] \:,
\end{split}
 \label{C_n eqn}	
\end{equation}
%\end{widetext}

\begin{equation}
%\begin{split}
C_p = C_n \left [\dfrac{1-\dfrac{3E_g}{k_B T}}{6\Bigg(1-\dfrac{5E_g}{4k_B T}\Bigg)} \right]\:,\\
% \end{split}
   \label{C_p eqn}	
\end{equation}

\begin{equation}
%\begin{split}
R_{A1} = \Big[p~ C_p + n~ C_n\Big]\:(np-n^2_i) \:,\\
% \end{split}
   \label{Auger recombiantion rate}	
\end{equation}

\noindent where $R_{A1}$ is the Auger 1 recombination rate. The overlap integral $|F_1 F_2|$ values range from 0.1 to 0.3. The key factor constraining the carrier lifetime in the IAS layers is the Auger 1 recombination. Given that holes are the sole carrier type capable of moving around the heterostructure, the intrinsic Auger 1 carrier lifetime $\tau^{A1}_i$ can be expressed as \cite{bellotti2006numerical,rogalski1985band}  
%\begin{widetext}
\begin{equation}
\begin{split}
\tau^{A1}_i=&3.8 \times10^{-18}~\dfrac{\epsilon^2_\infty\Big(1+\dfrac{m^*_e}{m^*_h}\Big)^{0.5} \Big(1+ 2~ \dfrac{m^*_e}{m^*_h}\Big)}{\Big(\dfrac{m^*_e}{m_0}\Big)~|F_1 F_2|^2}\times \\
&\Big(\dfrac{E_g}{k_BT} \Big)^{1.5} \exp \Bigg[\dfrac{\Big(1+2~ \dfrac{m^*_e}{m^*_h}\Big) E_g}{\Big(1+ ~ \dfrac{m^*_e}{m^*_h}\Big)k_B T} \Bigg] \:,\\
    \end{split}
	\label{intrinsic carrier lifetime}	
\end{equation}
%\end{widetext}

\begin{equation}
\tau^{A1}_h=\dfrac{2\tau^{A1}_i}{1+\Big( \dfrac{n_0}{n_i}\Big)^2}\:,
\label{holes carrier lifetime}
\end{equation}

\noindent where $\tau^{A1}_h$ is the hole carrier lifetime due to the Auger 1 recombination process, and $n_0$ is the equilibrium electron concentration. 

\section{Analysis of the dark current, optical responsivity, and the quantum efficiency}\label{various analysis}
The nBn architecture effectively mitigates the dark current contribution resulting from the SRH recombination. Consequently, the dark current in this architecture is mainly diffusion limited and can be calculated as \cite{martyniuk2013modeling,klipstein2008xbn,savich2015diffusion,kwan2021recent,martyniuk2014new,alchaar2019characterization} 
\begin{equation}
J_D = q\dfrac{n^2_i}{N_D}\dfrac{L_D}{\tau}\dfrac{\tanh \Big(\dfrac{t_{_{AL}}}{L_D}\Big)+\beta }{1+\beta ~ \tanh \Big(\dfrac{t_{_{AL}}}{L_D} \Big)}\:,
    	\label{dark current}	
\end{equation}
\noindent where $L_D$ is the diffusion length, $N_D$ is the donor density in the absorber region, $\tau$ is the minority carrier lifetime, and $t_{_{AL}}$ is the thickness of the AL. The surface recombination velocity, denoted as $\beta$, can be neglected due to the boundary conditions that enforce the absence of hole current at the interface between the AL and the CL. The valence band potential barrier resulting from the electrostatics of the junction has been taken into account. Only the thermally generated holes in the AL that possess sufficient kinetic energy to overcome the barrier will contribute to the dark current. Therefore, the dark current equation can be modified as 
\begin{equation}
J_D = q\dfrac{n^2_i}{N_D}\dfrac{L_D}{\tau} \tanh \Big(\dfrac{t_{_{AL}}}{L_D} \Big) \exp \Big[ \dfrac{E_a-3k_B T}{k_B T} \Big]\:.
    	\label{dark current modified}	
\end{equation}
The photocurrent density $J_{photo}$ caused by the incident power density $P_d$ was used to calculate the optical responsivity R as \cite{itsuno2011design,martyniuk2013modeling,shaveisi2023design,schuster2014analysis,kwan2021recent} 

\begin{equation}
R=\dfrac{J_{photo}}{P_d}\:.
    	\label{responsivity}	
\end{equation}

The quantum efficiency is a crucial parameter for assessing the optical response of the nBn detector and can be calculated as \cite{shaveisi2023design,martyniuk2013modeling,itsuno2011design,schuster2014analysis,kwan2021recent,akhavan2022design,martyniuk2013SPIE}

\begin{equation}
\eta=R\times \dfrac{h\nu}{q}\:.
    	\label{quantum efficiency_eq}	
\end{equation}

%\section*{References}
%\nocite{*}
%\bibliography{reference2}

%\section*{References}
%\nocite{*}
\bibliography{reference}
%\vspace{1.2cm}
\end{document}